\pdfoutput=1 
\documentclass[pre,notitlepage,groupedaddress,showkeys]{revtex4-1}

\usepackage[dvips]{graphicx}
\usepackage{amsmath,amssymb,bm}
\usepackage{color}
\usepackage{float}
\usepackage{subfig}
\begin{document}
\title{Dynamic behaviour of a ring coupled boost converter system with passivity-based control}

\author{Rutvika Manohar}
\email[]{r-manohar@dove.kuee.kyoto-u.ac.jp}
\author{Takashi Hikihara}
\email[]{hikihara.takashi.2n@kyoto-u.ac.jp}
\affiliation{Department of Electrical Engineering,
  Kyoto University,
  Katsura, Nishikyoku, Kyoto, Japan 615-8510 }

\date{\today}

\begin{abstract} 
This paper discusses a dispersed generation system of multiple  DC/DC converters with DC power sources connected in a ring formulation. Here is presented the analysis of the system based on the stored energy and passivity characteristics of the system. Passivity Based Control (PBC), with its energy-modifying and damping-injection technique, is applied to a ring coupled converter system to stabilize itself at a desired DC voltage in the presence of external disturbances.  The numerical results reveal the effective application of the control as a robust and flexible technique. 
\end{abstract}
\maketitle 
\section{Introduction}
\begin{figure}[H]
\centering
\includegraphics[width=0.4\hsize]{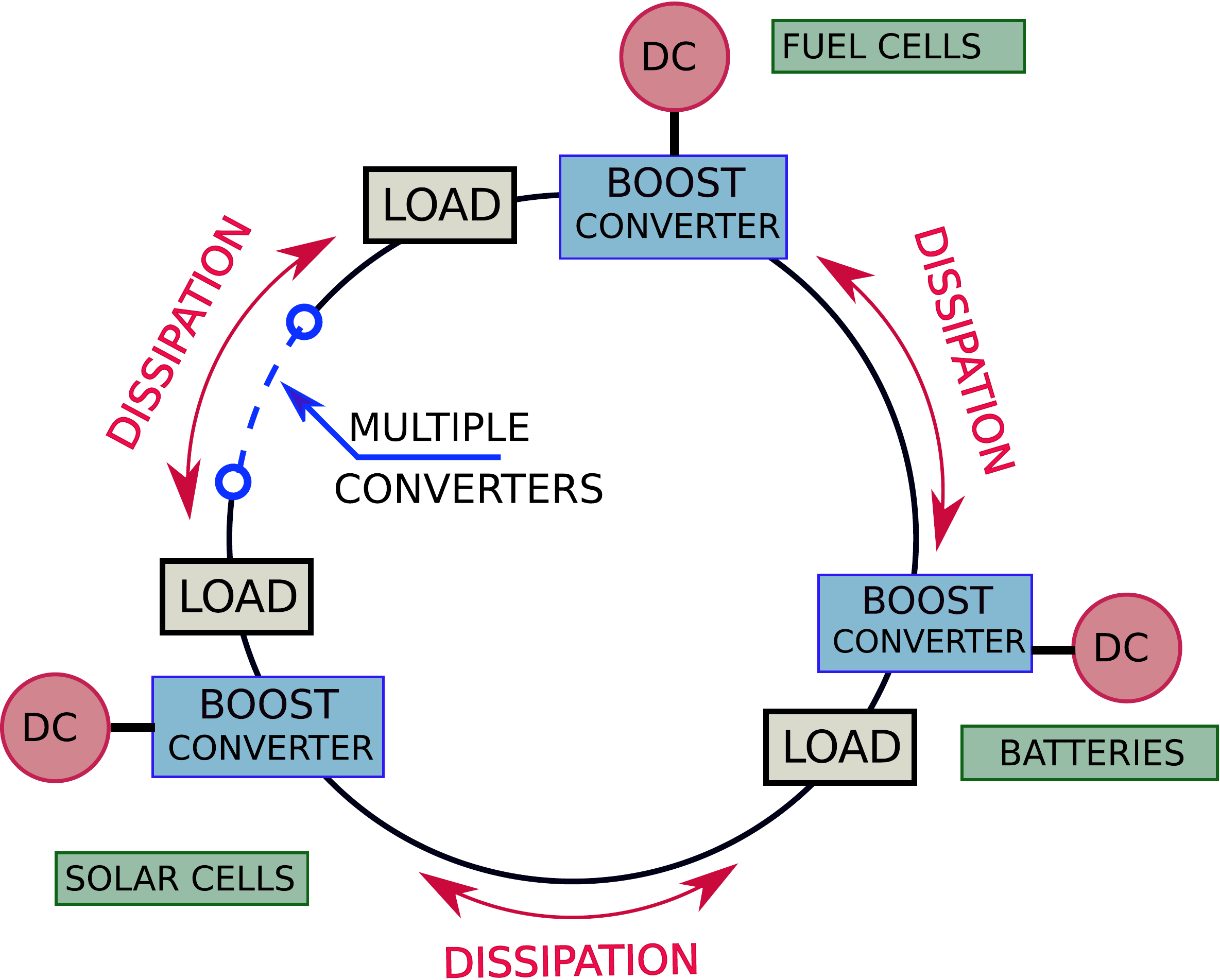}
\caption{System Configuration.}
\label{basic}
\end{figure}

Generation and storage of electricity with non-conventional power sources  like photovoltaic cells, fuel cells, wind power, and battries is realized with distributed generation \cite{dg-definition}. Power converters are seen as an important interface to transfer electricity from renewable sources of energy into the power network.  Towards designing a dispersed generation system, being autonomous from the conventional power grid, we propose a DC/DC converter system with DC power sources connected in a ring formulation and coupled with dissipation. The basic idea of such an `autonomous distributed generation system' is illustrated in Fig.\,\ref{basic}. This configuration is useful in harnessing energy from DC sources of electricity like solar cells, batteries, fuel cells and so on.
\par Passivity Based Control (PBC), aims at stabilizing the system by modifying the potential and kinetic energy function. It stems from the fundamental notion of `passivity', a property physically corresponding to the energy of the system.  Passive systems hold the characteristics in that the energy stored is always less than the energy supplied from outside \cite{passivity-dissipativity-springer}. PBC applied to passive systems is performed in two stages.  The first stage consists of shaping of the potential energy of the system such that the shaped potential energy function has a global and unique minimum at the desired equilibrium.  The second stage is to add damping to make the system exponentially stable \cite{ortega2013passivity}. 
\par The concept of passive systems was first developed in \cite{popov1973hyperstability} in the context of electrical circuits.  This, along with the framework of dissipative systems developed in \cite{willems1972dissipative}  helps to understand the concepts of passive systems and passivity based control from a dynamical systems point of view.
\par The energy-shaping plus damping-injection methodology used to solve state feedback set point regulation problems in fully actuated robotic systems by Takegaki and Arimoto, has evolved into `Passivity Based Control'  \cite{arimoto-manipulator}. The term `Passivity Based Control' was introduced first in \cite{ortega1989adaptive} to define a control methodology whose objective was to render the closed loop passive.  This control objective was first employed for the adaptive control of robot manipulators by \cite{landau1988synthesis}.  
\par Passivity Based Control has been traditionally applied to Euler-Lagrange systems \cite{ortega2013passivity}.  Here, the energy shaping stage accomplishes shaping of the potential energy and keeping the original kinetic energy to satisfy the passivation objective. For electronic systems which consists of non-energy elements like dissipation through resistance and switching elements, Lagrangian formulation is difficult or at times impossible to implement 
For systems like these, the kinetic as well as potential energy of the system need to be modified to design a storage function. Considering these resctrictions, instead of Euler-Lagrange modelling, Port-Controlled Hamiltonian modelling (PCHM) is used to achieve better results with PBC \cite{escobar1999hamiltonian}.  The modified storage function has to be shown to be a candidate of a Lyapunov function. 
\par The application of PBC to DC/DC converters was given in \cite{sira1995passivity}.  Then, the application to PBC for parallely connected DC/DC converters was discussed in \cite{hikihara2011regulation}. Building on the above mentioned works, this paper applies PBC to a ring coupled DC/DC converter system with multiple DC inputs. The DC/DC converter considered in this report is a boost converter, the characteristics of which have been well documented \cite{kassakian1991principles}. 
\par For ring-coupled converters, it is confirmed that the converters keep the property of passivity individually. In addition, the feedback control is employed with minimizing the energy function of the entire system, as opposed to that of a single DC/DC converter. The energy transfer between the individual converter units governs the the dynamic behaviour of the whole ring coupled system. The application of PBC, with `energy shaping' of the entire system, manifests the property of robustness despite the flow of energy between individual elements. 
\section{System Design and Modelling}
\subsection{System Equations}
We introduce a system consisting of multiple boost converters with multiple DC input sources.  These converters are coupled together with dissipation through inductance and resistance.  The coupling through dissipation represents a transmission line model, with inductive and resistive elements.   The basic configuration is shown in Fig.\,\ref{basic}. 

\par
\begin{figure}[H]
\centering
\includegraphics[width=0.6\hsize]{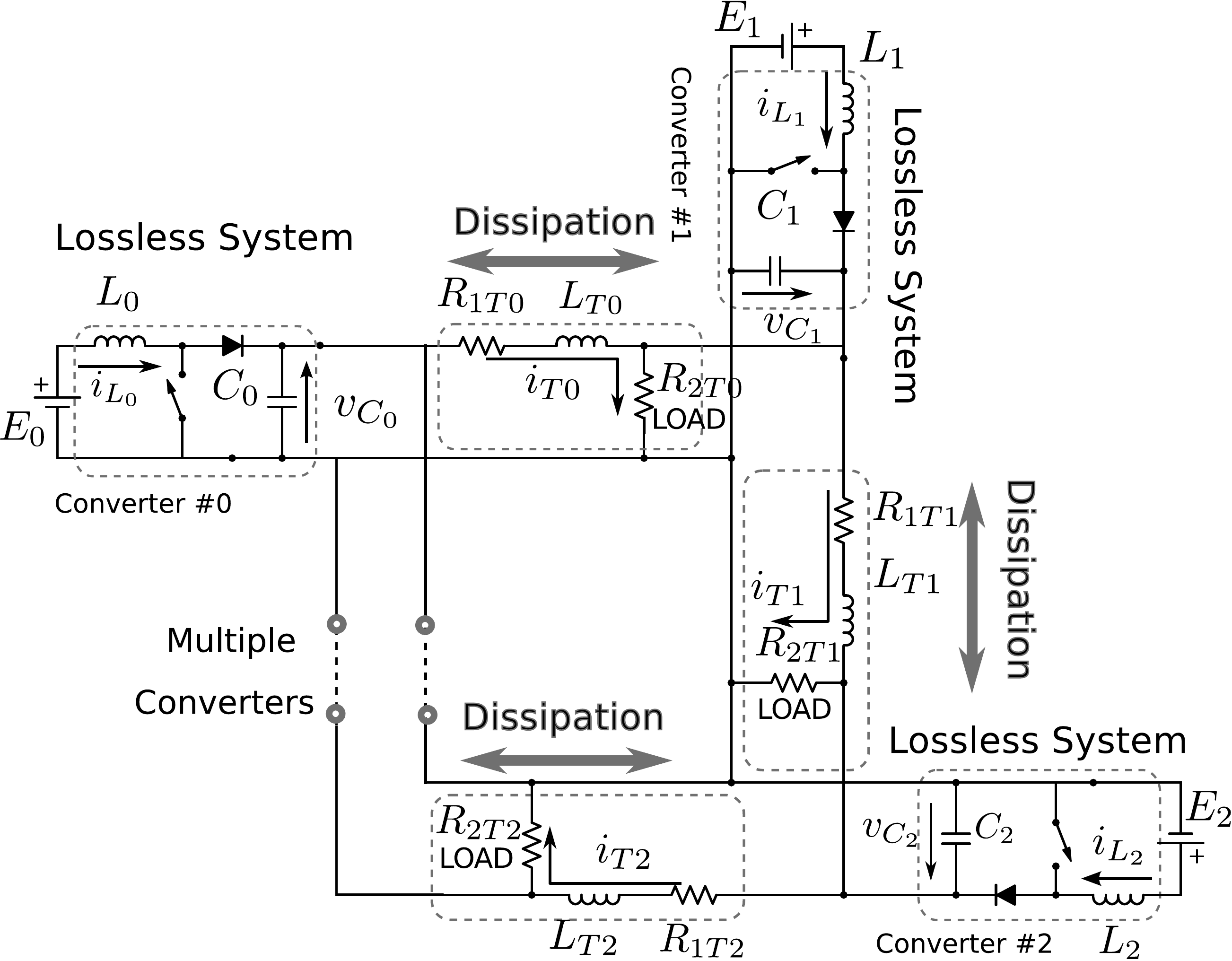}
\caption{Schematic diagram of ring coupled converters.}
\label{sys}
\end{figure}

 The schematic diagram  of the circuit is shown in Fig.\,\ref{sys}.  The coupled converters are responsible for the constant voltage output in the ring.  Here, load resistances($R_{2T}$) are across the output voltages whereas the dissipation elements of line inductor ($L_{t}$) and line resistance ($R_{1T}$) are in series.  The capacitor in the transmission line model is considered as negligible.  This is owing to the fact that a parallel  capacitor is dominant in the boost converter configuration.  The number of converters was set such that it enhances the asymmetry of the system.  The number of converters in the ring does not cause any loss of generality. Then, the objective is to apply PBC to the whole system, including the dissipation between the converters. 
 
\par The system equations are given by Eqs.(\ref{eqsys1})-(\ref{eqsys3}). Subscription $\mathrm{n}$ denotes the index of the converter. $u_{\mathrm{n}}$ denotes the switch position for the $\#\mathrm{n}$ converter. $u_{\mathrm{n}}=1$ implies that the switch is ON and $u_{\mathrm{n}}=0$ is OFF.  $L_{T\mathrm{n}}$ is the inductance and $R_{1T\mathrm{n}}$ in the dissipation between the \#$0$ and \#$1$ converter. Here, the dot notation represents differentiation with respect to time.

  \begin{align}
L_{\mathrm{n}}\dot{i}_{L\mathrm{n}}&= (1- u_{\mathrm{n}})v_{c\mathrm{n}} +E_{\mathrm{n}}\label{eqsys1}\\
C_{\mathrm{n}}\dot{v}_{C\mathrm{n}}&=(1-u_{\mathrm{n}})i_{L\mathrm{n}}-i_{T\mathrm{n}}+i_{T(\mathrm{n}-1)}-\frac{v_{c\mathrm{n}}}{R_{2T\mathrm{n}}} \label{eqsys2} \\
L_{T\mathrm{n}}\dot{i}_{T\mathrm{n}}&=v_{c\mathrm{n}}-v_{c(\mathrm{n}+1)}-R_{1T\mathrm{n}}i_{T\mathrm{n}}
  \label{eqsys3}  
  \end{align}
\subsection{Port Controlled Hamiltonian Modelling}
 Port Controlled Hamiltonian Modeling(PCHM) is the network representation of systems in interaction with their environment \cite{maschke1992port}. The Hamiltonian approach allows for the systematic modelling of electrical systems including resistors and switching elements.  The non-energetic terms are extracted from the circuit.  It leaves the energy conserving LC circuit.  The non-energy elements are then introduced into the circuit in the form of `ports'  \cite{maschke1992port}.  Such configuration of LC circuits with ports is represented as a generalized Hamiltonian system with external input variables.  PCHM technique is used for modeling DC/DC converters, to allow for the inclusion of power electronic switches and load resistances. 
\par  PCHM classifies the system neatly into physically well defined interconnection(\textbf{J}), dissipation(\textbf{R}) and external input(\textbf{E}) matrices within a state space framework \cite{ortega2013passivity}.  The system model using the Port Controlled Hamiltonian(PCH) framework is given by Eq.(\ref{pchs}) as in the form given in \cite{van2004port}.
\begin{align}
\begin{split}
\textbf{D}\bm{\dot{x}}(t)&=[\textbf{J}-\textbf{R}]  \frac{\partial H}{\partial x} +\textbf{E}\\ 
 \bm{y}&=g^{T}(\bm{x}) \frac{\partial H}{\partial \bm{x}}  
 \label{pchs}
\end{split}
\end{align}
\par For $\mathrm{m}$ converters in the configuration, $\bm{x}$, the state of the system, is a column matrix $((\mathrm{m}\times3) \times 1)$ of all the inductor currents and capacitance voltages. The matrix $\textbf{D}$ is a diagonal matrix of the capacitances and inductances of the corresponding currents and voltages.  $\textbf{J}$ gives the interconnection, and $\textbf{R}$ the dissipation in the system. The interconnection is a function of $u_{\mathrm{n}}$, the switch position of the corresponding boost converters in the ring configuration. $\textbf{E}$ , the input matrix is a column matrix of the input voltages to the respective converters. $H$ is the Hamiltonian, which, in this case, is the total energy of the system. 
\begin{equation}
\\
	\\
\\
\bm{\mathrm{D}}=
\begin{bmatrix}
\textbf{D}_0  & 0 & \dots & 0\\
0         &\textbf{D}_1 & \dots &0\\
\vdots & \vdots & \ddots & \vdots\\
0 & 0 & \dots & \textbf{D}_\mathrm{m}\\
\end{bmatrix}
\text{where , }
\textbf{D}_{\mathrm{n}}
  \begin{bmatrix}
   L_{\mathrm{n}} & 0&0\\
   0 & C_{\mathrm{n}}&0 \\
   0 & 0&L_{T\mathrm{n}} \\
    \end{bmatrix}.
    \label{Dmatrix}
\end{equation}
 
\begin{equation}
	\bm{x}=
    \begin{bmatrix}
 \bm{x}_{0}\\
\vdots\\
\bm{x}_{\mathrm{m}}
  \end{bmatrix}
  \text{where , }
  \bm{x}_{\mathrm{n}}=
\begin{bmatrix}
i_{L\mathrm{n}}\\
v_{C\mathrm{n}}\\
i_{T\mathrm{n}}\\
\end{bmatrix}.\\ \\
\label{xmatrix}
\end{equation}
\\
\begin{equation}
	\bm{\mathrm{J}}=
  \begin{bmatrix}
\textbf{A}_0&-\textbf{B}^T&0&\dots&0&\textbf{B}\\
\textbf{B}&\textbf{A}_{1}&-\textbf{B}^T&0&\dots&0\\
0&\textbf{B}&\ddots&\ddots&\ddots&\vdots\\
\vdots&0&\ddots&\ddots&\ddots&0\\
0&\vdots&\ddots&\ddots&\ddots&-\textbf{B}^T\\
-\textbf{B}^T&0&\dots&0&\textbf{B}&\textbf{A}_{\mathrm{m}}\\
\end{bmatrix}\\ 
\label{Jmatrix}
\end{equation}
\\
\begin{equation}
\text{Here, }
\\
\textbf{A}_\mathrm{n}=
\begin{bmatrix}
0&-(1-u_{\mathrm{n}})&0\\
(1-u_{\mathrm{n}})&0&-1\\
0&1&0\\
\end{bmatrix},
\textbf{B}=
\begin{bmatrix}
0&0&0\\
0&0&1\\
0&0&0\\
\end{bmatrix}.
\label{Bmatrix}
\end{equation}
\\
\begin{equation}
	\\ 
\textbf{R}=
\begin{bmatrix}
\textbf{R}_0  & 0 & \dots & 0\\
0 &\textbf{R}_1 & \dots&0\\
\vdots & \vdots & \ddots & \vdots\\
0 & 0 & \dots & \textbf{R}_\mathrm{m}\\
\end{bmatrix}
\text{, where }
\textbf{R}_{\mathrm{n}}=
  \begin{bmatrix}
  0 & 0&0\\
   0 &\cfrac{1}{R_{2T\mathrm{n}}}&0 \\
   0 & 0&R_{1T\mathrm{n}} \\
    \end{bmatrix}.\\
    \label{Rmatrix}
    \end{equation}
\\ \\
Looking at the adjacency matrix of the interconnection matrix (\textbf{J}), the ring coupled structure is clearly verified. It is seen that $\textbf{D}$ is a diagonal matrix and $\textbf{R}$ is a symmetric matrix for $\mathrm{m}$ converters. The matrix $\textbf{J}$ is the interconnection matrix, and shows the coupling between neighbouring converters through dissipation.  From Eq.(\ref{Jmatrix}), it can be verified that $\textbf{J}$ is a skew-symmetric matrix, with keeping the PCHM structure \cite{van2004port}.

\section{Application of Passivity-Based Control } 
The control objective for the ring coupled converter system is to regulate the output voltage of the ring towards a desired equilibrium value. First we analyse the steady state behaviour of the system and then stabilize the system with a  feedback control.
\subsection{Steady State Equations of the System}
  Till now we have considered the switched model for the boost converters \cite{kassakian1991principles,sira2006control}. The control objective is to regulate the average output voltage to a constant reference (e.g DC/DC converters) or a periodic value with frequency much smaller than the switching frequency (inverters). Thus, it desirable to consider the average value of the voltages and currents rather than the instantaneous values, given that the ripple and harmonics are sufficiently small. 
  \par The averaging of the circuit implies the averaging of the circuit variables as well as the switching function which gives the switch position. In order to obtain the averaged model of the switching function, we replace the switch position $u_{\mathrm{n}}$ in the instantaneous circuit with a modulating signal $\mu_{\mathrm{n}}$ for the averaged circuit.  A pulse width modulation (PWM) policy is implemented for switch regulation for the converters.  PWM is a method for generating a digital pulse signal to drive the transistor switches in the boost converter circuit from a continuous control input or a modulating signal.  If the modulating signal varies slowly in comparison to the switching frequency, it can be shown that the switching function is equal to the modulating signal \cite{kassakian1991principles}.   As all the other components in the circuit are linearly time independent (LTI), they can be just replaced by their average values with any other treatment unchanged. The only modification necessary is the replacement of the switching function with the modulating signal for the PWM, which will be referred to as the duty ratio. Here after, in this paper, the state variables $i_{L\mathrm{n}}$, $i_{T\mathrm{n}}$, $v_{C\mathrm{n}}$ will be considered the average values of the instantaneous variables, and $\mu_{\mathrm{n}}$ as the duty ratio. 
  \par To gain insight into the steady state behaviour of the system, it is desirable to establish the relationship between the equilibrium values of the average output voltage, the average input current, and the average dissipation current. Towards this end a constant duty ratio $\mu_{\mathrm{n}}=U_{\mathrm{n}}$ is implemented. Then, it is clear from Eqs.(\ref{eqsys1})-(\ref{eqsys3}) that the equilibrium values of the state variables are given by Eqs.(\ref{ss1})-(\ref{ss3}).
  \setlength{\jot}{15pt}
\begin{align}
\begin{split} \label{ss1}
\bar i_{L\mathrm{n}} ={}&\frac{E_{\mathrm{n}}}{R_{2T_{\mathrm{n-1}}}(1-U_{\mathrm{n}})^{2}}+\frac{1}{R_{1T_{\mathrm{n}}}}\bigg[\frac{E_{\mathrm{n}}}{(1-U_{\mathrm{n}})^{2}}-\frac{E_{\mathrm{n+1}}}{(1-U_{\mathrm{n+1}})(1-U_{\mathrm{n}})}\bigg ]\\
&-\frac{1}{R_{1T_{\mathrm{n-1}}}}\bigg[\frac{E_{\mathrm{n-1}}}{(1-U_{\mathrm{n-1}})(1-U_{\mathrm{n}})}-\frac{E_{\mathrm{n}}}{(1-U_{\mathrm{n}})^{2}}\bigg]
\end{split}\\
\bar v_{C\mathrm{n}}={}&\frac{E_{\mathrm{n}}}{(1-U_{\mathrm{n}})} \label{ss2}\\ 
\bar i_{T_{\mathrm{n}}}={}&\frac{1}{R_{1T_{\mathrm{n}}}}\bigg[\frac{E_{\mathrm{n}}}{(1-U_{\mathrm{n}})}-\frac{E_{\mathrm{n+1}}}{(1-U_{\mathrm{n+1}})}\bigg] \label{ss3}
\end{align}
\\
  
   
 Here, $\bar i_{L\mathrm{n}} $ denotes steady state inductor current, $\bar v_{C\mathrm{n}}$ the steady state output voltage, and $\bar i_{T_{\mathrm{n}}}$ the steady state inductor current for converter \#n.
\par The desired output voltage $ v_{C\mathrm{n}d}$ decides the duty ratio. In other words,  the duty ratio is used to keep the output in the ring at $ v_{C\mathrm{n}d}$.  This is an open loop system.  The next section gives the estimation of the open loop system for the comparing to the results with closed loop system with PBC.
\subsection{Consideration of Zero Dynamics}
In this section we consider the analysis of the averaged ring coupled converter system through to the zero dynamics at equilibrium points. Zero dynamics are defined as a dynamics that characterizes the internal behaviour of the system once the initial conditions and inputs are chosen such that the output is identically zero \cite{isidori2013zero}. Here, we consider the `zero dynamics' associated with the equilibrium points given for the output as the capacitor voltage and the inductor current respectively.
\par Firstly, the output voltage of the capacitor is regarded as the output of the averaged PWM model of the ring coupled system. Then, rewriting the equations given by Eqs.(\ref{eqsys1})-(\ref{eqsys3}) in terms of $v_{C\mathrm{n}}$ the following relation is obtained.

\begin{align}
C\ddot{v}_{C}&=-\dot{\mu}_{\mathrm{n}} i_{L_{\mathrm{n}}}+(1-\mu_{\mathrm{n}})\dot{i}_{L_{\mathrm{n}}}-\dot{i}_{T\mathrm{n}}+\dot{i}_{T\mathrm{n}}-\frac{v_{C\mathrm{n}}}{R_{2T\mathrm{n}}}\\
&=\frac{-\dot{\mu}_{\mathrm{n}}}{(1-\mu)}\bigg[C\dot{v}_{C\mathrm{n}}+i_{T\mathrm{n}}-i_{T\mathrm{n-1}}+\frac{v_{C\mathrm{n}}}{R_{1\mathrm{n}}}\bigg]+\frac{(1-\mu_{\mathrm{n}})}{L}\bigg[E_{\mathrm{n}}-(1-\mu_{\mathrm{n}})\bigg]-(\dot{i}_{T\mathrm{n}}-\dot{i}_{T\mathrm{n-1}})-\frac{\dot{v}_{C\mathrm{n}}}{R_{2\mathrm{n}}}
\end{align}
The objective for the zero dynamics is to chose the control variable $\mu_{\mathrm{n}}$ so as to keep the output voltage constrained at $v_{C\mathrm{n}}=v_{Cd}$ and $i_{T\mathrm{n}}=\bar{i}_{T\mathrm{n}}$. Then, it can be conferred that $\ddot{v}_{C}=0$ and $\dot{v}_{C\mathrm{n}}=0$. Thus, the output is fixed but the control is not and hence $\dot{\mu_{\mathrm{n}}} \neq 0$.  Then, the following relation is held. 
\begin{equation}
\dot{\mu}_{\mathrm{n}}=\frac{(1-\mu_{\mathrm{n}})^2}{L[(\bar{i}_{T\mathrm{n}}-\bar{i}_{T\mathrm{n-1}})+\bar{v}_{C\mathrm{n}}]}\bigg[E-(1-\mu)\bar{v}_{C\mathrm{n}}\bigg]
\label{dotmu}
\end{equation}

\par
\begin{figure}[H]
\centering
\includegraphics[width=0.5\hsize]{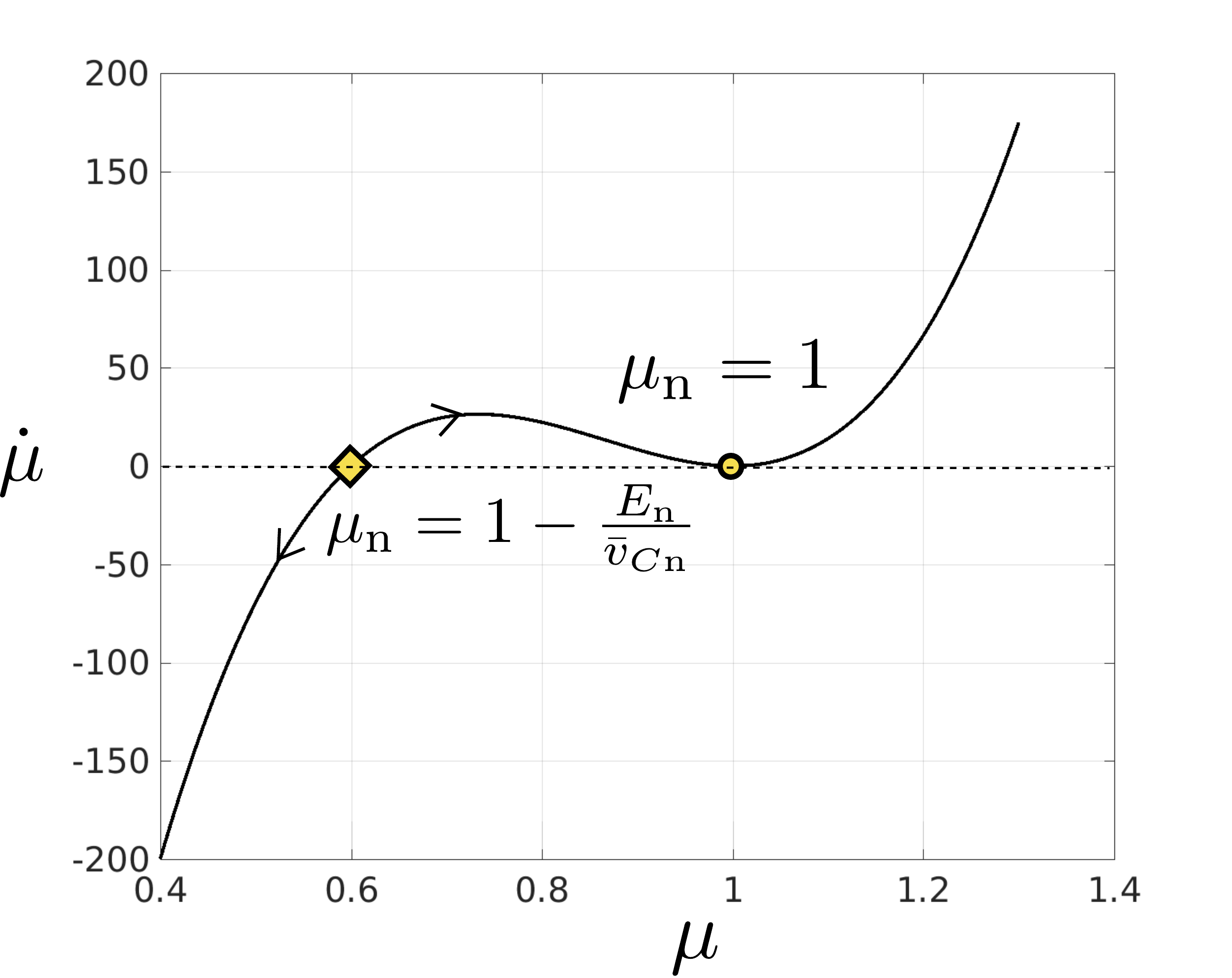}
\caption{Zero dynamics of ring coupled converter system corresponding average output capacitor voltage.}
\label{zerodyn1}
\end{figure}

 The equilibrium points of Eq.(\ref{dotmu}) are given at $\mu_{\mathrm{n}}=1-E_{\mathrm{n}}/\bar{v}_{C\mathrm{n}}$ and $\mu_\mathrm{n}=1$. Among these, the first has phisical significance. If $\bar{v}_{C\mathrm{n}}>E_{\mathrm{n}}$, it is confirmed that the output will be always higher than the input of the converter. However, the phase diagram describes this equilibrium point as unstable, making the system non-minimum phase with respect to the output voltage. The phase diagram is drawn in accordance with the parameters given in Table.\,\ref{tablbal} and is given by Fig.\,\ref{zerodyn1}. Such analysis is performed for single boost converter in \cite{sira2006control} and \cite{ortega2013passivity}.

Next, the analysis of the system corresponding to the zero dynamics with the inductor current as the output of the averaged PWM model is carried out. Then, rewriting the equations given by Eqs.(\ref{eqsys1})-(\ref{eqsys3}) in terms of $i_{l\mathrm{n}}$ the following equation is obtained.
\begin{equation}
L_{\mathrm{n}}\ddot{i}_{L\mathrm{n}}=\dot{\mu}\bigg[\frac{E_{\mathrm{n}}-L_{\mathrm{n}}\dot{i}_{L\mathrm{n}}}{(1-\mu)}\bigg]-\frac{(1-\mu)}{L_{\mathrm{n}}C_{\mathrm{n}}}\bigg[(1-\mu)i_{l\mathrm{n}}-i_{T\mathrm{n}}+i_{T\mathrm{n-1}}-\frac{E_{\mathrm{n}}-L\dot{i}_{L\mathrm{n}}}{(1-\mu)R_{2\mathrm{n}}}\bigg]
\label{sysil}
\end{equation}
As same as before, the control variable $\mu_{\mathrm{n}}$ can be chosen so as to keep the output at a constant value $i_{l\mathrm{n}}=\bar{i}_{L\mathrm{n}}$ and $i_{T\mathrm{n}}=\bar{i}_{T\mathrm{n}}$. Then, it follows that $\ddot{i}_{L\mathrm{n}}=0$ and $\dot{i}_{L\mathrm{n}}=0$. The remaining dynamics related to the control variable, i.e the duty ratio $\mu_{\mathrm{n}}$ are described by Eq.(\ref{dotmui}).
\begin{equation}
\dot{\mu}=\frac{(1-\mu)}{R_{\mathrm{2n}}C_{\mathrm{n}}E_{\mathrm{n}}}[R_{\mathrm{2n}}(1-\mu)^2\bar{i}_{L\mathrm{n}}-R_{\mathrm{2n}}(1-\mu)(i_{T\mathrm{n}}-i_{T\mathrm{n}})-E_{\mathrm{n}}]
\label{dotmui}
\end{equation}
The three equilibrium points corresponding to Eq.(\ref{dotmui}) are obtained as Eq.(\ref{muil}). 

\begin{align}
\mu_{\mathrm{n,1}}&=1-\frac{(\bar{i}_{T\mathrm{n}}-\bar{i}_{T\mathrm{n-1}})}{2\bar{i}_{L\mathrm{n}}} - \sqrt{\frac{E_{\mathrm{n}}}{R_{\mathrm{2n}}}-\bigg[\frac{(\bar{i}_{T\mathrm{n}}-\bar{i}_{T\mathrm{n-1}})}{2\bar{i}_{L\mathrm{n}}}\bigg]}\nonumber\\
\mu_{\mathrm{n,2}}&=1-\frac{(\bar{i}_{T\mathrm{n}}-\bar{i}_{T\mathrm{n-1}})}{2\bar{i}_{L\mathrm{n}}} + \sqrt{\frac{E_{\mathrm{n}}}{R_{\mathrm{2n}}}-\bigg[\frac{(\bar{i}_{T\mathrm{n}}-\bar{i}_{T\mathrm{n-1}})}{2\bar{i}_{L\mathrm{n}}}\bigg]}\nonumber\\
\mu_{\mathrm{n,3}}&=1
\label{muil}
\end{align}

\par
\begin{figure}[H]
\centering
\includegraphics[width=0.5\hsize]{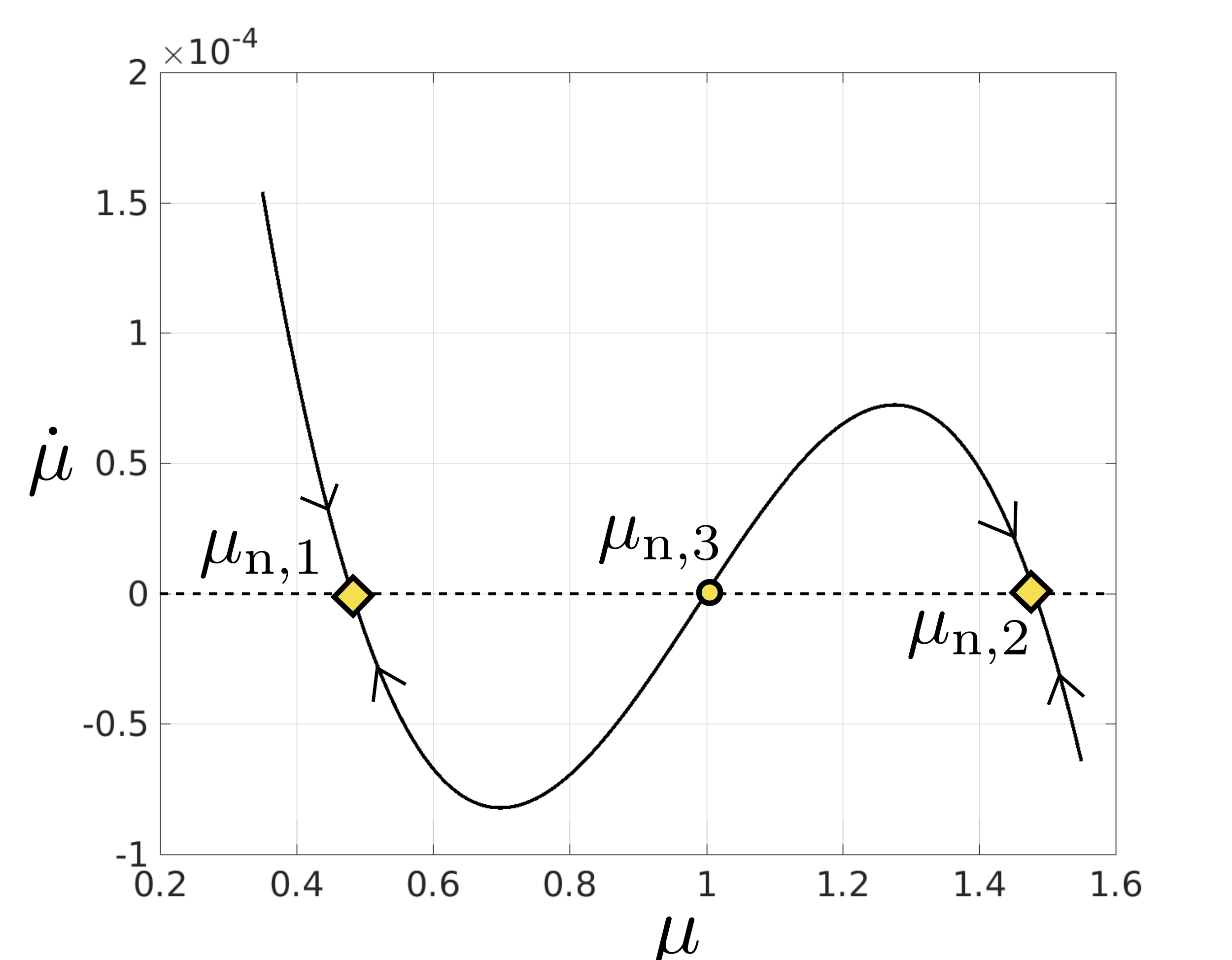}
\caption{Zero dynamics of ring coupled converter system corresponding average output inductor current.}
\label{zerodyn2}
\end{figure}

 As $(\bar{i}_{T\mathrm{n}}-\bar{i}_{T\mathrm{n-1}})$ is sufficiently small, we can safely neglect it the squared terms. It can also be established that $(\bar{i}_{T\mathrm{n}}-\bar{i}_{T\mathrm{n-1}})/2\bar{i}_{L\mathrm{n}}<< \sqrt{E_{\mathrm{n}}/R_{\mathrm{2n}}}$. Thus, only one equilibrium point satisfies the condition of $0\leq\mu_{\mathrm{n}}\leq 1$ and is of physical significance. All the equilibrium points are given by Fig.\,\ref{zerodyn2}, in accordance with the parameters specified in Table.\,\ref{tablbal} . It is seen that the equilibrium point is stable, as long as $Ri_{Ld}>E$.  This again emphasizes the properties of the boost converter, making the system controlled with inductor current a minimum phase system.  The inductor current will be used to stabilize the system rather than the output capacitor voltage. 
\subsection{Energy Shaping in PBC}\label{energyshapingpbc}
We investigate whether it is possible for PBC to be applied to a system of multiple converters coupled with dissipation. 
 The stored energy of a circuit is the sum of the energy in the passive elements, that is the inductors and capacitors.  The energy of the multiple converter system is given by Eq.(\ref{hamiltonianringcoupled}).
\begin{equation}
	H= \frac{1}{2}\bm{x}^T\textbf{D}\bm{x}
	\label{hamiltonianringcoupled}
\end{equation}
$\bm{x}$ and $\textbf{D}$ are as given in Eqs.(\ref{xmatrix}) and (\ref{Dmatrix}), respectively.  $H$, which is the total energy, is also considered to be the Hamiltonian of the system. 
 \par The desired states can be formed based on the Hamiltonian as discussed in Chap.2. The formulation is taken to be the quadratic function of errors.  It is given by Eq.(\ref{newdesirederror}).
 \begin{equation}
 	H_{d}=\frac{1}{2}\bm{e}^T\textbf{D}\bm{e},\,\,\,\,\, \bm{e}=\bm{x}-\bm{x}_{d}
 	\label{newdesirederror}
 \end{equation}
Here, $\bm{x_d}$ is the desired trajectory of the state.  In order to prove that this function is a candidate of Lyapunov function,  let us check the derivative of the energy function as in Eq.(\ref{derlyapunovsys}).
\begin{equation}
\begin{split}
\dot H_{d}&= (\bm{x}-\bm{x}_{d}(t))^{T}\textbf{D}(\bm{\dot x} - \bm{\dot x}_{d}(t))\\
&=(\bm{x}-\bm{x}_{d}(t))^{T}([\textbf{J}-\textbf{R}]\bm{x}+ \textbf{E}- \textbf{D}\bm{\dot x}_{d}(t))
\end{split}
\label{derlyapunovsys}
\end{equation}
By setting the term $\textbf{D}\bm{\dot{x}}^{*}(t)$ as given in Eq.(\ref{boostexo}),  the control rule with PBC can be formulated for boost converters.
\begin{equation}
\textbf{D}\bm{\dot{x}}^{*}(t)=(\textbf{J}-\textbf{R})\bm{x}^{*}(t)+\textbf{E}+\textbf{R}_{I}(\bm{x}-\bm{x}^{*}(t))
\label{boostexo}
\end{equation}
Where, $\textbf{R}_{I}$ is a symmetric and positively defined matrix, which acts as the damping injection, making the system asymptotically stable if $\textbf{R}+\textbf{R}_{I}>0$. By considering that $\bm{e}^T\textbf{J}(\mu)\bm{e}=0$ for all values of $\mu$, we get the following condition on the derivative of our chosen Lyapunov function $H_{d}(\bm{e})$.
\begin{equation}
\dot{H}_{d}(e)=\bm{e}^T(\textbf{J}(\mu)\bm{e}-\textbf{R}\bm{e}-\textbf{R}_{I}\bm{e}=-\bm{e}^T(\textbf{R}+\textbf{R}_{I})\bm{e} < 0
\label{boostlyapunovfinal}
\end{equation}
\par Under the control rule, the error $\bm{e}$ converges to the origin asymptotically.  Then,  by satisfying the condition $(\textbf{R}+\textbf{R}_{I})>0$, the system becomes exponentially and asymptotically stable at the equilibrium point \cite{khalil1996noninear}. Then the derivative of $H_{d}(e)$ is given in Eq.(\ref{boostlyapunovfinalder}).
\begin{equation}
\dot{H}_{d}(\bm{e})=-\bm{e}^T(\textbf{R}+\textbf{R}_{I})\bm{e} \leq -kH_{d}(\bm{e})
\label{boostlyapunovfinalder}
\end{equation}
Hence it is proved that Lyapunov's Theorem is satisfied. That is, the equilibrium state $\bm{x}^*$ is asymptotically stable with the control rule given in Eq.(\ref{boostexo}), and exponentially asymptotically stable with the damping injection. $\textbf{R}_I$ is given in Eq.(\ref{RImatrix}).
\begin{equation}
	\\ 
\bm{\mathrm{R}}=
\begin{bmatrix}
\bm{\mathrm{R}_{I0}}  & 0 & \dots & 0\\
0 &\bm{\mathrm{R}_{I1}} & \dots&0\\
\vdots & \vdots & \ddots & \vdots\\
0 & 0 & \dots & \bm{\mathrm{R}_{I\mathrm{n}}}\\
\end{bmatrix}
\text{, where }
\bm{\mathrm{R}_{\mathrm{In}}}=
  \begin{bmatrix}
  R_{\alpha\mathrm{n}} & 0&0\\
   0 &0&0 \\
   0 & 0& 0\\
    \end{bmatrix}.\\
    \label{RImatrix}
    \end{equation}
    \setlength{\jot}{15pt}
\begin{align}
\dot{H}_{d}(e)&=\bm{e}^T(\textbf{J}(\mu))\bm{e}-\textbf{R}\bm{e}-\textbf{R}_{I}\bm{e}\\
&=-\bm{e}^T(\textbf{R}+\textbf{R}_{I})\bm{e}<0
\label{boostlyapunovfinal}
\end{align}
Then, the control equation is obtained for the given system in Eq.(\ref{dutycontrol}).  The boost converter is controlled with inductor current for keeping passivity.  The constant desired inductor current ($i_{Ld}$) is obtained by solving the steady state Eq.(\ref{ss1}).
\setlength{\jot}{12pt}
	\begin{align}
	\mu_{\mathrm{n}}&= \frac{1}{{v}_{C\mathrm{n}d}}\big[E_{\mathrm{n}}+R_{I \mathrm{n}}(i_{L\mathrm{n}}-\bar{i}_{L\mathrm{n}})\big]+1 \label{dutycontrol} \\
	C_{\mathrm{n}}\dot{{v}}_{C\mathrm{n}d} &=(1-\mu_{\mathrm{n}})  \bar{i}_{L\mathrm{n}}-  i_{T_{\mathrm{n}d}}+  i_{T_{\mathrm{n-1}d}}- \big[\frac{ v_{C\mathrm{n}d}}{R_{2T_{\mathrm{n-1}}}}\big]\\
L_{T_{\mathrm{n}}}\dot{{i}}_{T_{\mathrm{n}d}} &=  v_{C\mathrm{n}d}-  v_{C\mathrm{n+1}d}- R_{1T_{\mathrm{n}}} i_{T_{\mathrm{n}d}}\label{itcontrol}
\end{align}

\par The value of the duty ratio $\mu$ is evaluated at every instant $t$ depending on the input, the parameters, and the desired output of the system. That is, $\mu$ depends on time and the state. Hereafter, when PBC is applied to the system, $\mu$ is considered as a function of time.

\section{Numerical Simulations}
The simulation results are obtained for five converters coupled in a ring form are given in this section. The numerical simulations were carried out on ode45 solver Simulink (Version 8.7 R2016a).
\begin{table}[htb]
\parbox{.45\linewidth}{
\centering
\caption{Parameters for Balanced System}\label{tablbal}
\vspace{2mm}
\begin{tabular}{llll}
\hline
Parameter& Value &Unit\\
\hline
$E_{\mathrm{n}}$  &  15   &\textrm{V}\\
\hline
 $L_{\mathrm{n}}$  &   46  &\textrm{mH}\\
\hline
$C_{\mathrm{n}}$&  100 & \textrm{$\mu$F}  \\ 
\hline
 $L_{T\mathrm{n}}$ &15 & \textrm{mH}\\ 
\hline
 $R_{1T\mathrm{n}}$ & 100&$\Omega$ \\ 
\hline
$R_{2T\mathrm{n}}$ & 170 & $\Omega$\\ 
\hline
$v_{Cd\mathrm{n}}$&   40& \textrm{V}   \\ 

\end{tabular}
}
\hfill
\parbox{.45\linewidth}{
\centering
\caption{Different Input Voltages}\label{tablunbalbal1}
\vspace{2mm}
\begin{tabular}{llll}
 \hline
Parameter& Value &Unit\\
\hline
$E_{\mathrm{0-4}}$  &  15,13,12,13,15   &\textrm{V}\\
\hline
 $L_{\mathrm{n}}$  &   46  &\textrm{mH}\\
\hline
$C_{\mathrm{n}}$&  100 & \textrm{$\mu$F}  \\ 
\hline
 $L_{T\mathrm{n}}$ &15 & \textrm{mH}\\ 
\hline
 $R_{1T\mathrm{n}}$ & 100&$\Omega$ \\ 
\hline
$R_{2T\mathrm{n}}$ & 170 & $\Omega$\\ 
\hline
$v_{Cd\mathrm{n}}$&   40& \textrm{V}   \\ 
\end{tabular}

}
\end{table}
\begin{table}[htb]
\begin{center}
\caption{Different Load Resistances}\label{tablunbalbal4}
\fontsize{9}{12}\selectfont
\begin{tabular}{llll}
 \hline
Parameter& Value &Unit\\
\hline
$E_{\mathrm{0-4}}$  &  15 &\textrm{V}\\
\hline
 $L_{\mathrm{n}}$  &   46  &\textrm{mH}\\
\hline
$C_{\mathrm{n}}$&  100 & \textrm{$\mu$F}  \\ 
\hline
 $L_{T\mathrm{n}}$ &15 & \textrm{mH}\\ 
\hline
 $R_{1T\mathrm{n}}$ & 100&$\Omega$ \\ 
\hline
$R_{2T\mathrm{0-4}}$ & 130,170,140,170,130 & $\Omega$\\ 
\hline
$v_{Cd\mathrm{n}}$&   40& \textrm{V}   \\ 
\end{tabular}
\end{center}
\end{table}
\begin{figure}[H]
\centering
\includegraphics[width=0.7\hsize]{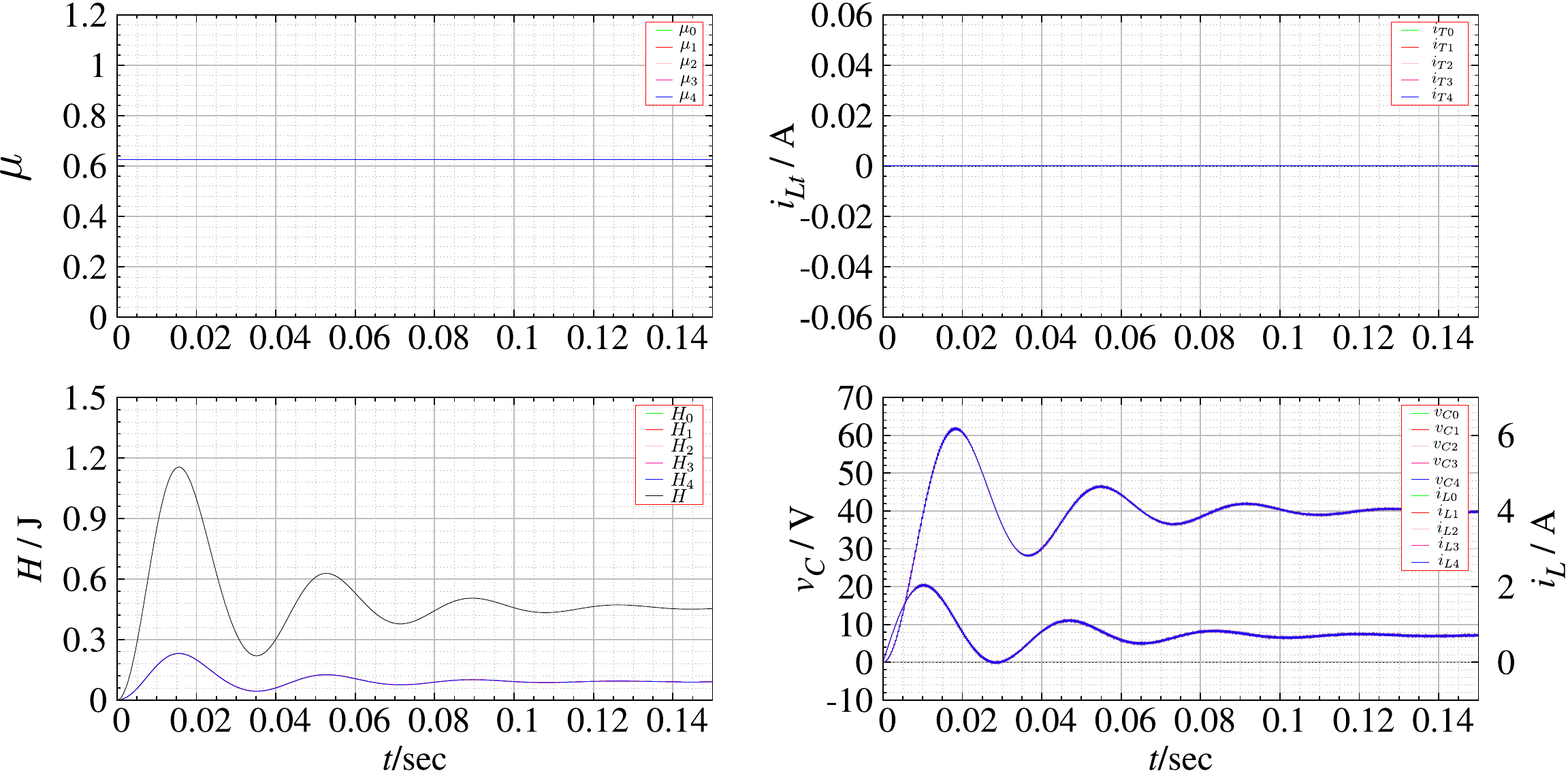}
\caption{Balanced system behaviour of original system with respect to time. Parameters of all converters are set at equal values. The desired output voltage  $v_{Cd}=40$ $\mathrm{V}$ for all $\mathrm{5}$ converters. The output of all converters coincides through the transient and settles at $40$ $\mathrm{V}$.}
\label{nonconbal}
\end{figure}
\subsection{Simulation results for a balanced system }\label{simulationwithoutcontrol}
\par Parameters in the ring coupled converter system are set at the same values respectively. This is a balanced case.  Additionally, the dissipation between the converters ($L_{T\mathrm{n}}, R_{1T\mathrm{n}}$) as well as the input $E_{\mathrm{n}}$ are also set at same values for all the converters. This naturally creates an energy balance in the ring.  To begin with, numerical simulations are performed for a balanced system without applying any feedback control.  Then, the duty cycle was set at a constant value ($U$). This value is calculated by Eq.(\ref{ss2}) for a particular $v_{Cd}$.    Thus, any disturbance in the system will not be acted on but will affect the system output adversely.  The parameters for simulation are given in Table.\,{\ref{tablbal}}.  
Fig.\,\ref{nonconbal}  shows the inductor voltage, dissipation current, output voltage and duty cycle with respect to time.  In the transient, the output voltage and the inductor current oscillate.  After the transient, the system reaches an equilibrium. The equilibrium value of the output voltage becomes equal to $v_{Cd}$.  As all the boost converters have same inputs and same parameter values, no current flows through the dissipation.

\par
\begin{figure}[H]
\centering
\includegraphics[width=0.7\hsize]{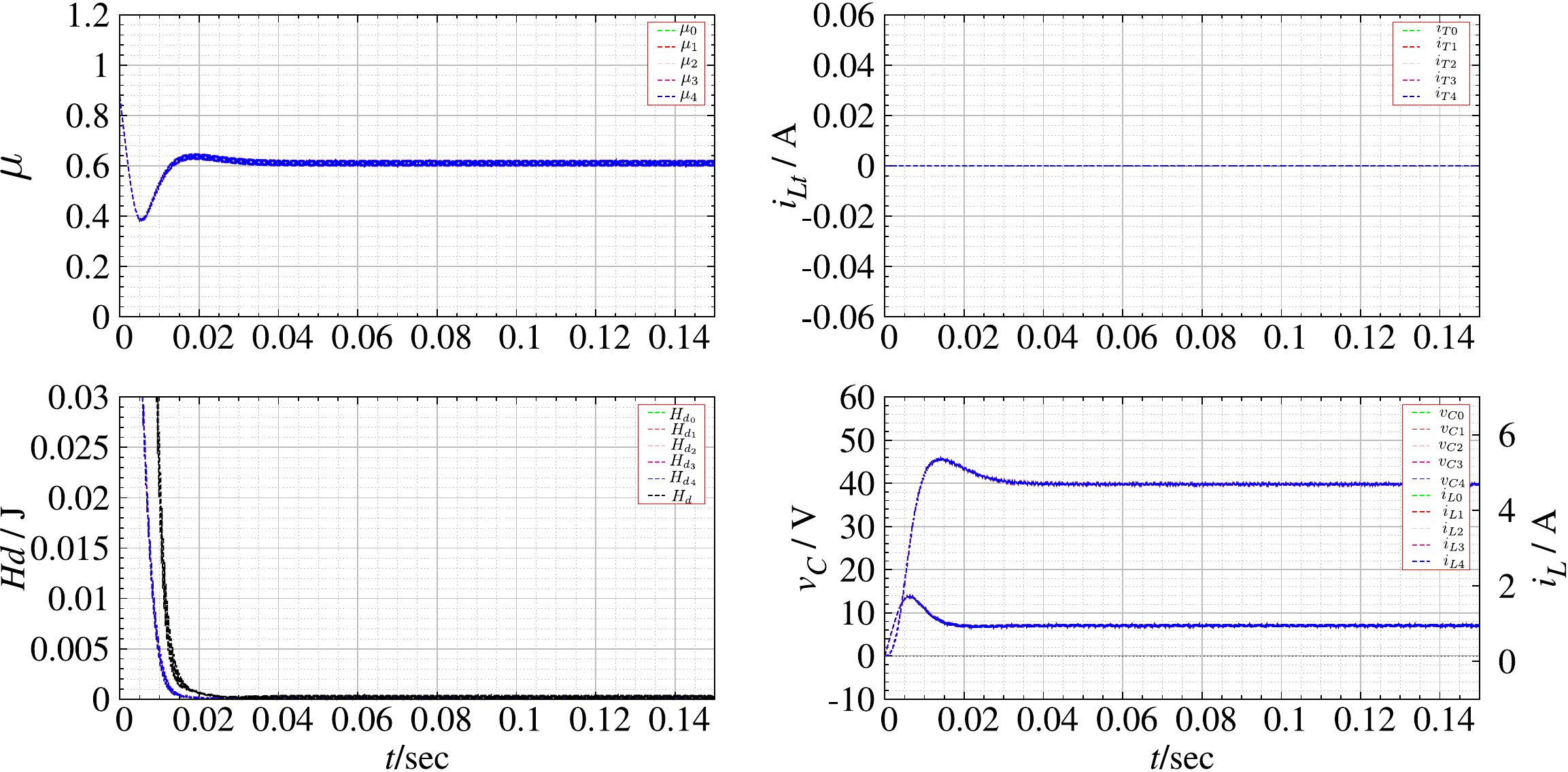}
\caption{Balanced system behaviour with application of PBC to all converters. The energy function $H_{d}$ goes to zero as system settles to the desired equilibrium. As the system is balanced, all the converters move synchronously. The transient peak is damped and the convergence time is improved.}
\label{conbal}
\end{figure}

 Next we look at simulations results when PBC is applied. PBC results in an output dynamic feedback controller which induces a shaped closed loop energy and enhances the closed loop damping of the system.  The damping has the condition $\textbf{R}+\textbf{R}_{I}>0$.  Then it was set and kept at $\textbf{R}_{I1}=15$ for all the simulations. The results, for a balanced case, show faster response and damped oscillations.  Feedback is provided by solving Eq.(\ref{dutycontrol}) to obtain the appropriate duty cycle value to maintain the desired equilibrium voltage.  Here, the inductor current takes a constant value as it is the variable that is measured to provide feedback. The numerical simulation results are shown in Fig.\,\ref{conbal}
\begin{figure}[H]
\centering
\includegraphics[width=0.5\hsize]{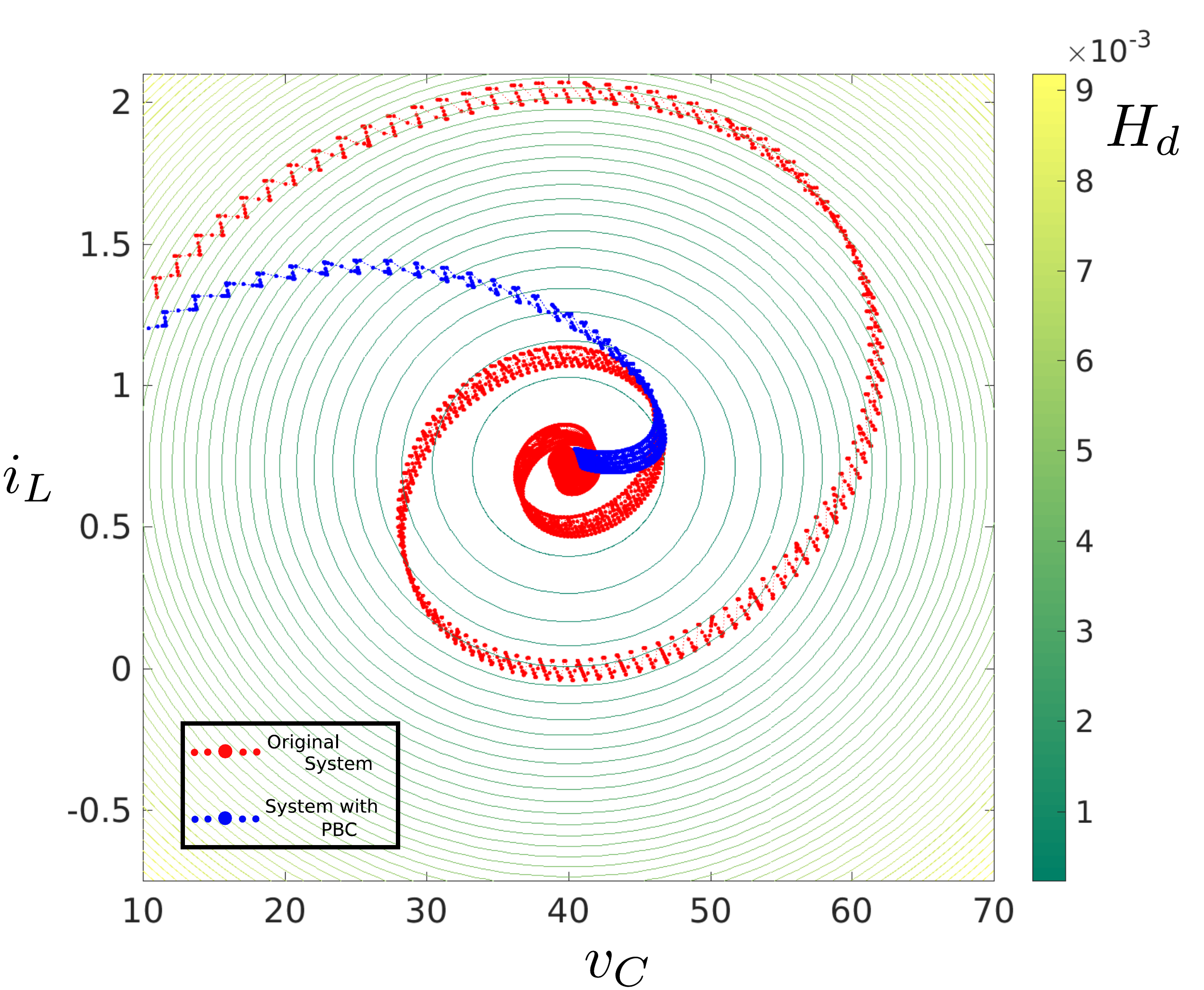}
\caption{Asymptotic behaviour of the original system and system with PBC}
\label{asymptoticbehaviour}
\end{figure}

\par The simulation parameters set as given in Table.\,\ref{tablbal}.  Comparing Fig.\,\ref{conbal} with Fig.\,\ref{nonconbal}, it is seen that the transient is sufficiently dampened.  The time to stabilize the system to the equilibrium is also significantly faster.  As all the parameters are the same for all boost converter systems as well as the dissipation, the trajectories on the output voltage and inductor current plane coincide for all five converters. 
The asymptotic behaviour is clearly observed with an phase plot on the energy sets of $H_{d}$. This is shown in Fig.\,\ref{asymptoticbehaviour}.

\subsection{Simulation results for an unbalanced system }

\par Imbalance occurs in the ring coupled system when converters in the ring and/or the dissipation between two neighbouring converters have different parameter values.  In the following simulations we consider imbalance in two different ways: imbalance created with varying input voltage and load resistance values.  First, simulations were carried out to see the behaviour of the original unbalanced system without application of PBC. The values of the input voltages are varied as given in Table.\,\ref{tablunbalbal1}. 

\par
\begin{figure}[H]
\centering
\includegraphics[width=0.7\hsize]{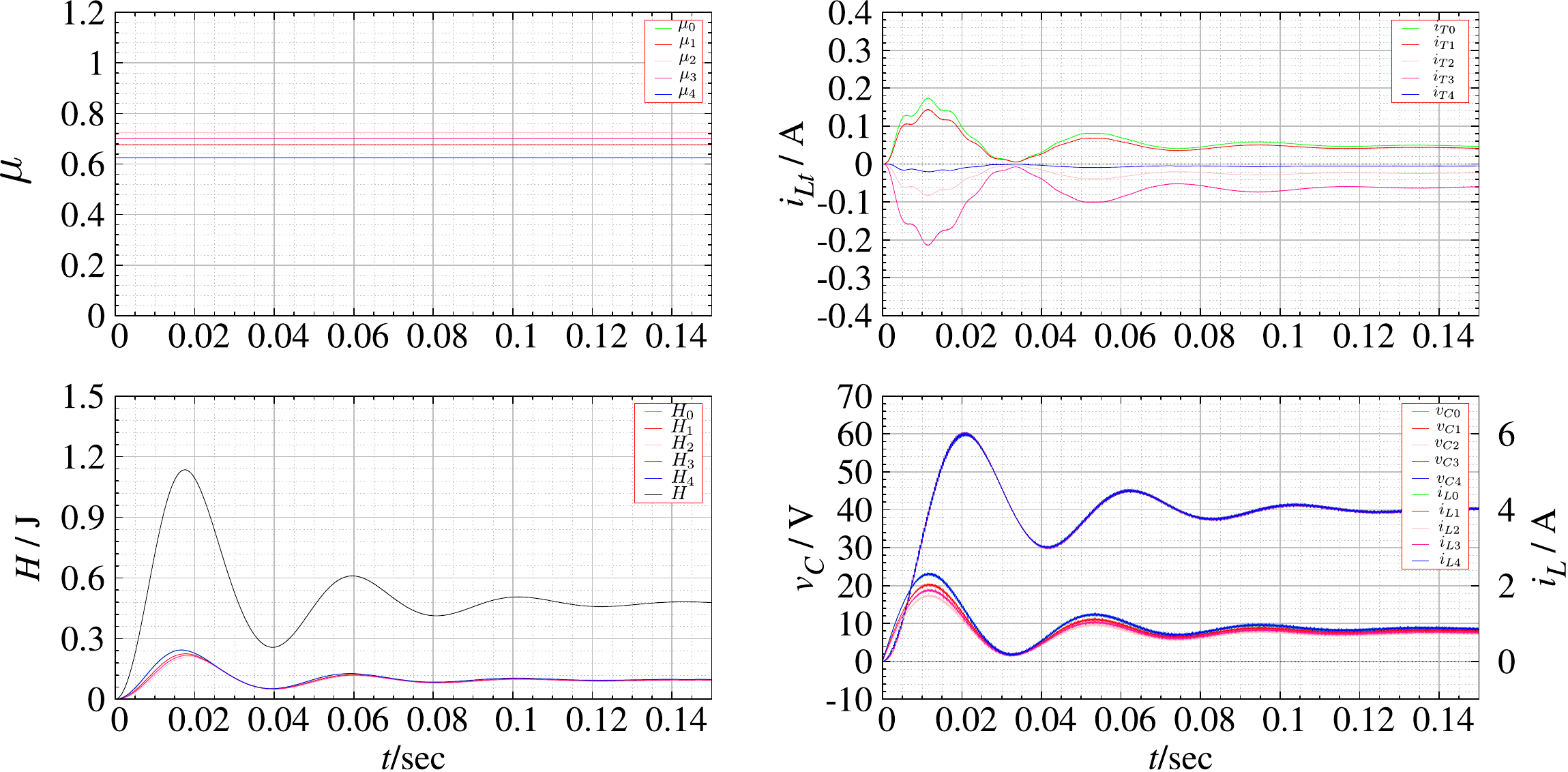}
\caption{Original unbalanced system with different input voltages to the boost converter. $E_{0}=15$ $\mathrm{V}$, $E_{1}=13$ $\mathrm{V}$, $E_{2}=12$ $\mathrm{V}$, $E_{3}=13$ $\mathrm{V}$, $E_{4}=15$ $\mathrm{V}$.}
\label{nonconunbal1}
\end{figure}

 Fig.\,\ref{nonconunbal1} shows that different input voltages give rise to different current values for the same desired voltage.  In this case, all converters have constant but different duty cycles as calculated from Eq.(\ref{ss2}).  Due to the different inputs, it is found that current appears through the dissipation between the coupled converters. The dissipation current is reduced after the transient when the converters converge to the desired voltage. 

\par
\begin{figure}[H]
\centering
\includegraphics[width=0.7\hsize]{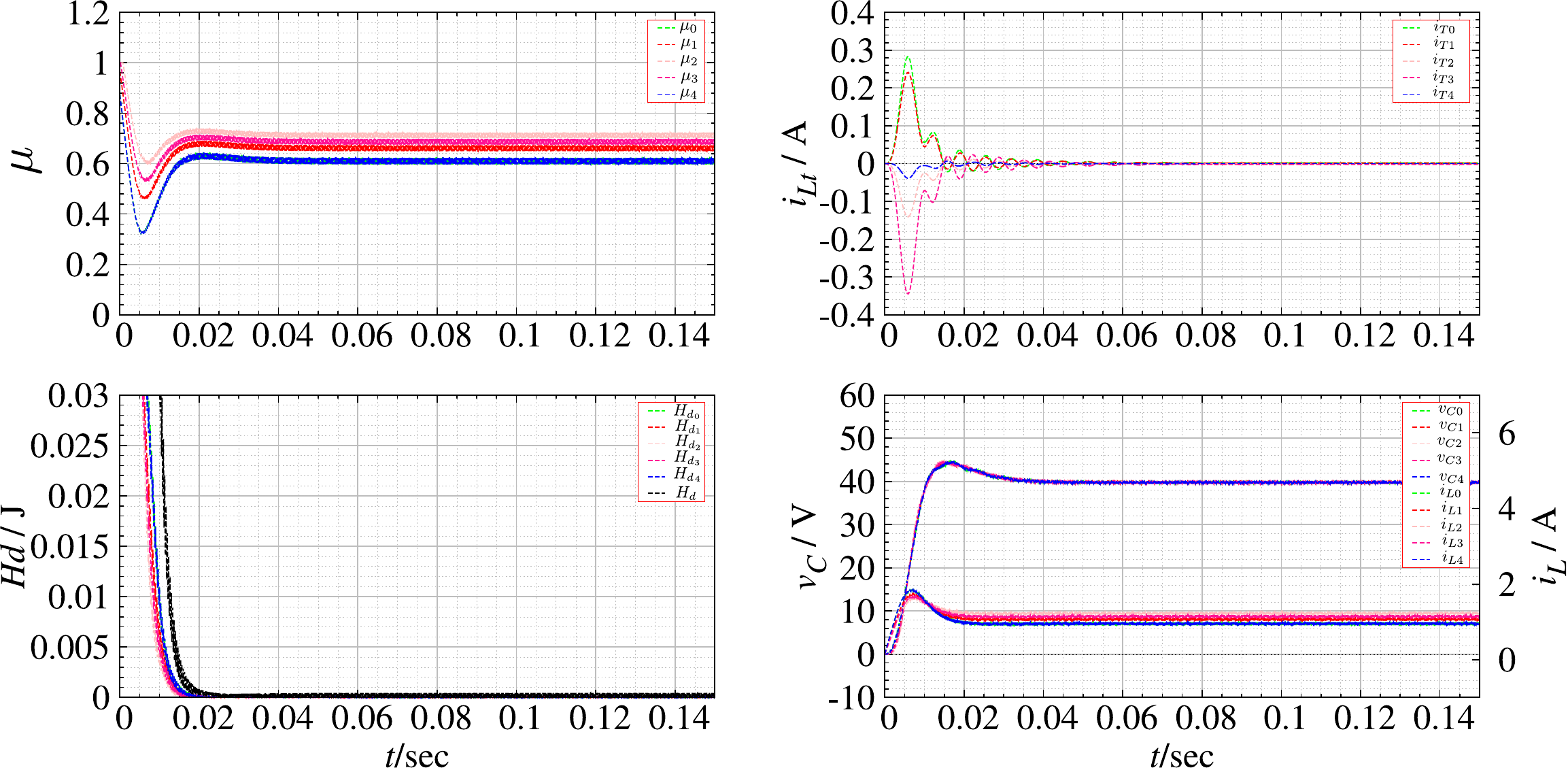}
\caption{Unbalanced system with different input voltages for the boost converters with PBC. $E_{0}=15$ $\mathrm{V}$, $E_{1}=13$ $\mathrm{V}$, $E_{2}=12$ $\mathrm{V}$, $E_{3}=13$ $\mathrm{V}$, $E_{4}=15$ $\mathrm{V}$.}
\label{conunbal1}
\end{figure}

  Figure \ref{conunbal1} shows simulation results when PBC is applied.  The direction of flow of the dissipation current depends on the imbalance created by the different inputs.  The duty ratio changes according to the energy function ($H_{d}$) and becomes constant as soon as the energy function attains a zero value. The energy function for each of the converter is different for the unbalanced case. It is seen in the results that each energy function becomes zero as the control is applied. This implies that PBC is successfully applied to each converter system as well. Thus, it can be confirmed that interconnection of passive systems is a passive system \cite{passivity-dissipativity-springer}. Even though energy is exchanged between converter systems during transient period, passivity is retained for each converter, and all converters stabilize at desired equilibrium. 
\par
\begin{figure}[H]
\centering
\includegraphics[width=0.7\hsize]{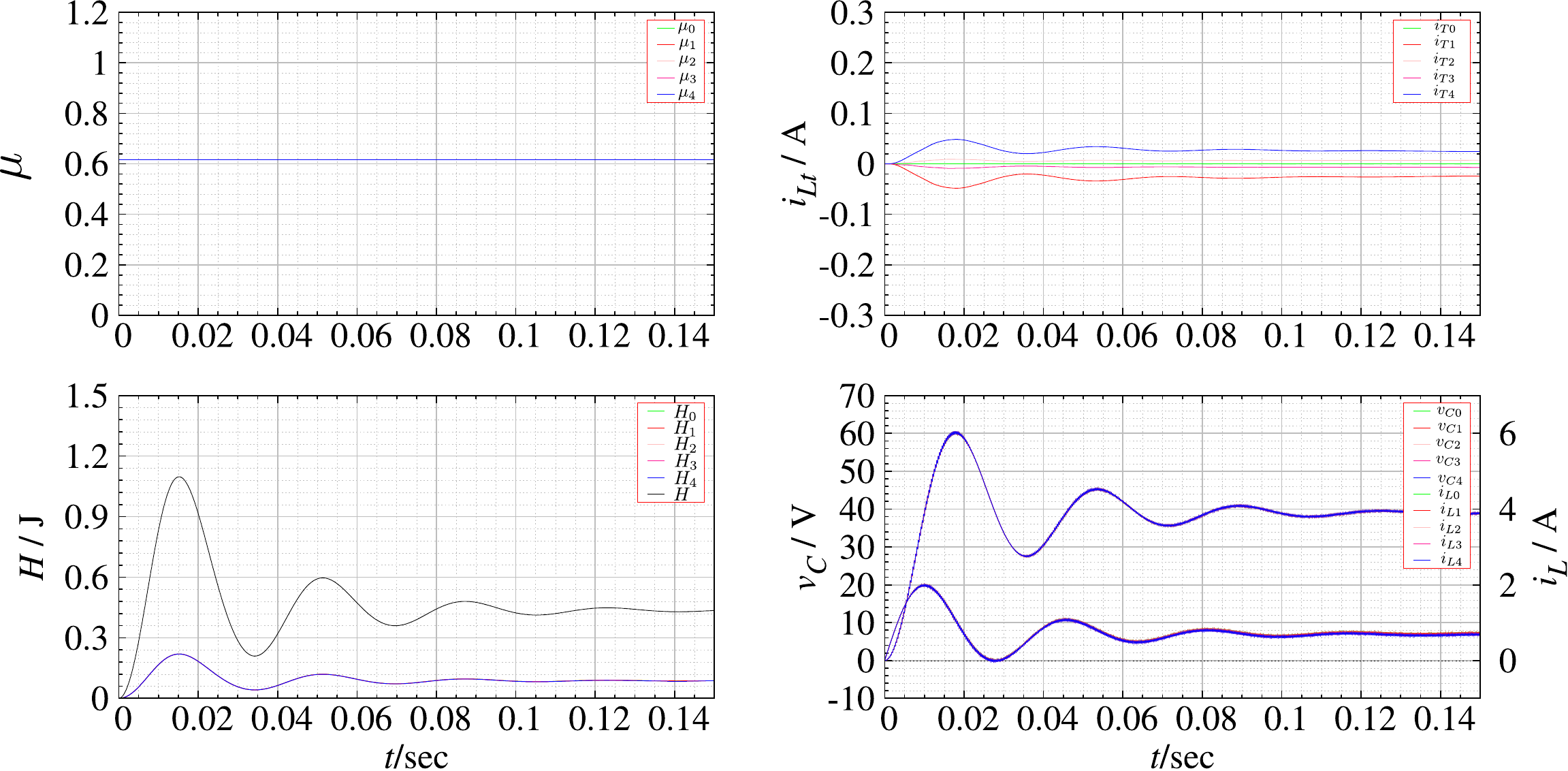}
\caption{Original system with imbalance created with different load resistances $R_{2T0}=30$ $\Omega$, $R_{2T1}=170$ $\Omega$, $R_{2T2}=140$ $\Omega$, $R_{2T3}=170$ $\Omega$, $R_{2T4}=130$ $\Omega$.}
\label{nonconunbal4}
\end{figure}

 Next, the load resistance is set differently for at each of the converters as given in Table.\,\ref{tablunbalbal4}.  The simulation results for the original system are shown in Fig.\,\ref{nonconunbal4}. The different load resistances cause the voltages to settle at slightly different values and this causes the dissipation current to keep flowing between the converter systems. This indicates the flow of energy between the converters. The flow of energy is in the direction of the load that consumes the most current, i.e towards the smallest resistance. 
\begin{figure}[H]
\centering
\includegraphics[width=0.7\hsize]{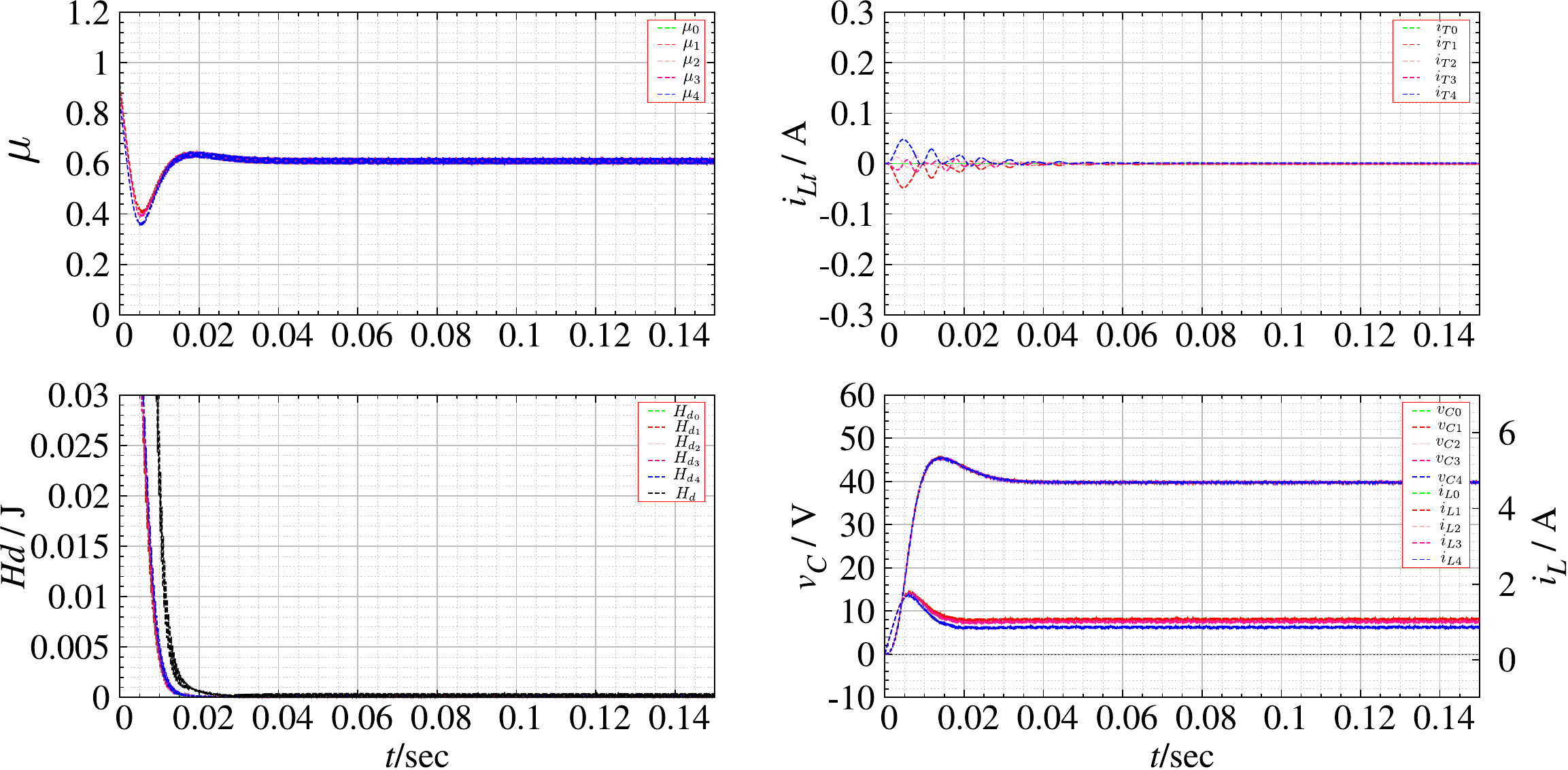}
\caption{Unbalanced system with different load resistances with PBC. $R_{2T0}=30$ $\Omega$, $R_{2T1}=170$ $\Omega$, $R_{2T2}=140$ $\Omega$, $R_{2T3}=170$ $\Omega$, $R_{2T4}=130$ $\Omega$.}
\label{conunbal4}
\end{figure}

\par Finally, Fig.\,\ref{conunbal4} shows the results of PBC applied to the system when the load resistance values are different. As is expected, the results are better than the uncontrolled case. This implies that PBC can successfully regulate the coupled converter systems when the system has different loads. The loads in a realistic system would vary depending on the power requested by the user. Therefore, the successful stabilization of system with PBC with different loads serves as an useful tool for designing practical systems.

\subsection{Stabilization to non stationary state}
\begin{figure}[H]
\centering
\includegraphics[width=0.5\hsize]{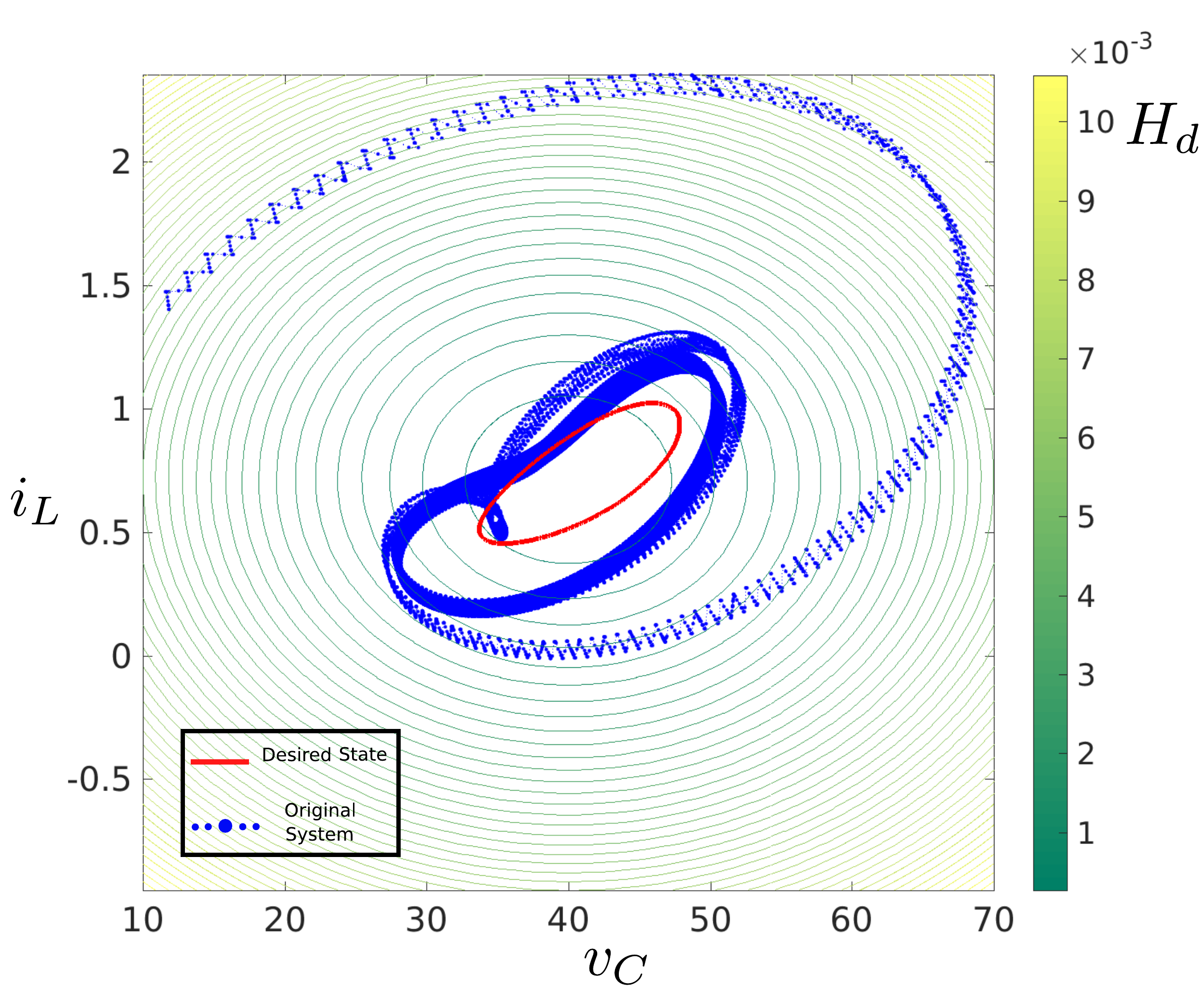}
\caption{Asymptotic behaviour of open loop system with the desired equilibrium as a function of time}
\label{nocontrolsineasymp}
\end{figure}

So far the control has been constructed assuming that the desired state, $v_{Cd}$ is a constant.  We may further extend PBC to the case in which the desired state is not a constant but a function on time. A sinusoidal function with a DC bias is selected as the desired state. The desired state is set to be $v_{Cd}=v_{DC}+A\sin(2 \pi f t)$. Here, $v_{DC}$ is the DC bias voltage, $A$ and $f$  are the amplitude and frequency of the sinusoidal voltage. The frequency is much smaller than the switching frequency of the PWM. For simulations we set $v_{DC}=40$ $\mathrm{V}$, $A=8$ $\mathrm{V}$, $f=60$ $\mathrm{Hz}$.  
Firstly, the simulations were carried out for the original system, where the duty ratio $U_{\mathrm{n}}$ is calculated from Eq.(\ref{ss2}). The results are given in the form of an energy plot (Fig.\,\ref{nocontrolsineasymp}). The other parameters are set as given by Table.\,\ref{tablbal}.

\par
\begin{figure}[H]
\centering
\includegraphics[width=0.5\hsize]{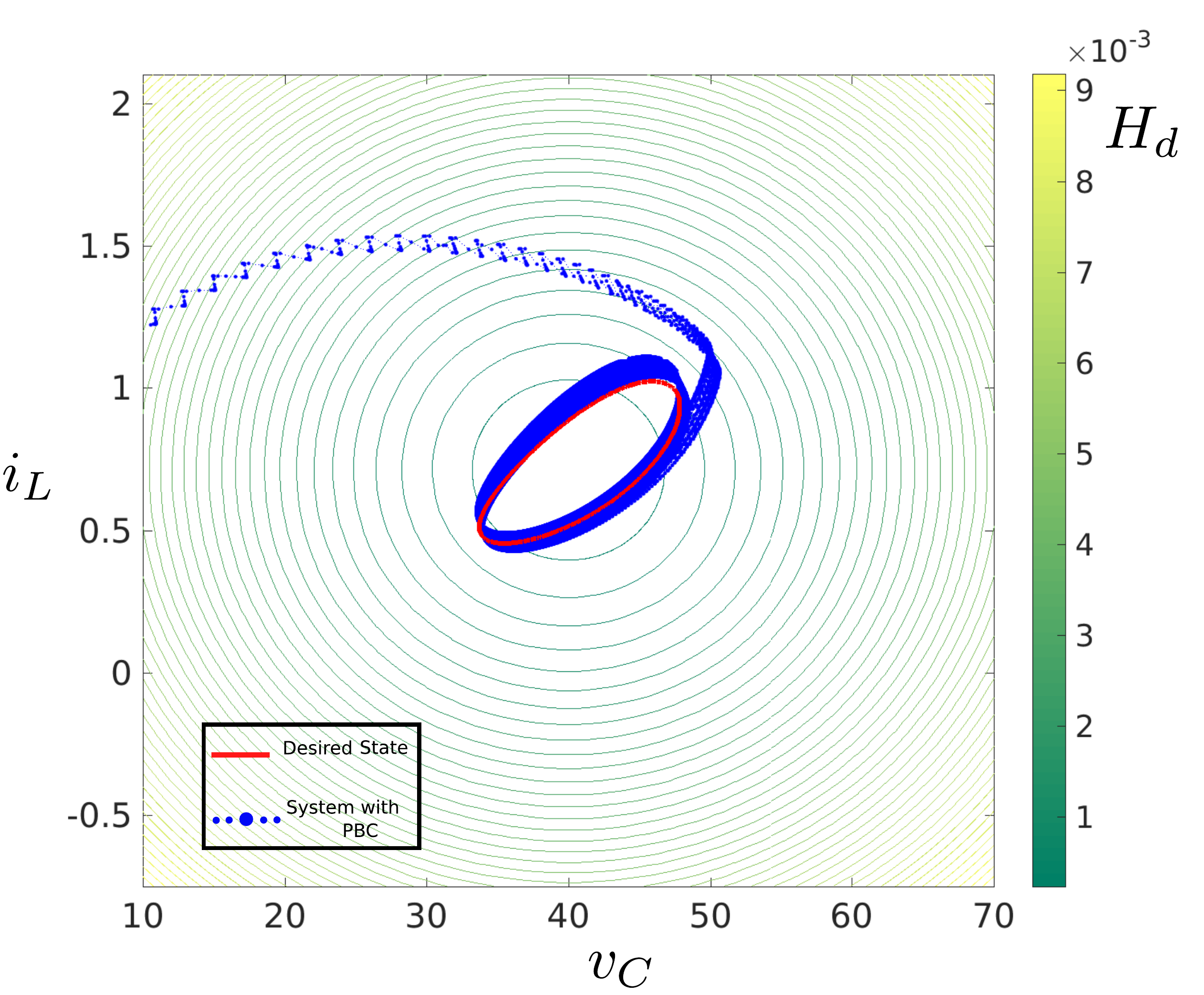}
\caption{Asymptotic behaviour with application of PBC having desired equilibrium as a function of time}
\label{controlsineasymp}
\end{figure}

 Next, simulations were carried out by applying PBC as given by Eqs.(\ref{dutycontrol}) and (\ref{itcontrol}). The results are given in Fig.\,\ref{controlsineasymp}.  Again, the parameters are set as given by Table.\,\ref{tablbal}.


\par It was seen in Sec.\,\ref{energyshapingpbc}, that the desired function is considered as a function of time. The subsequently obtained control equations for the feedback through the duty cycle also emphasize the fact that it is permissible to have the desired state as a function of time. It implies that PBC is a suitable method of control for such a case. The simulation results confirm the improvement at the application of PBC. The open loop system exhibits an unstable transient and a high transient peak voltage. The system with control settles down to the desired equilibrium relatively faster.

\section{Conclusion}
In this paper we proposed a method to stabilize a ring coupled converter system, consisting of DC/DC boost converters, to a desired state with the application of PBC. 
 PBC, with energy shaping and damping injection was discussed for the quadratic function of errors as the desired storage function. The desired storage function deviates around zero and finally approaches zero as the system attains equilibrium. 
\par Numerical simulations show that PBC can stabilize the output voltage values at the desired state even for a ring coupled configuration of the DC/DC converters. Comparison to the dynamic behaviour of the original system suggests the successful application of PBC during transient operation. Numerical simulations were carried for different initial conditions.
This included a balanced condition, where all the boost converters have same parameters, an unbalanced state with different parameters, including the dissipation, and the case in which the desired state is non stationary. PBC was applied for all the three cases. The results for the balanced system show all converters in the ring operating synchronously. There is no energy exchange in the form of dissipation current. Imbalance causes energy imbalance, but the application of PBC restores this imbalance and the entire system stabilizes at the desired state. Practically, the input voltages to the converters as well as the load resistances are not same for all converters in the ring. Even in such imbalanced conditions, the converters co-operate to maintain a stable voltage through the ring coupling. For non stationary desired states, the convergence of the output voltage as well as the inductor current to the desired sinusoidal state is vastly improved under PBC as compared to the original system, with a dampened transient peak.

\par 

\bibliographystyle{ieeetr}
\bibliography{manohar} 
\end{document}